\documentclass[journal=nalefd,10pt,manuscript=letter,layout=twocolumn]{achemso}   	
\usepackage{bm}
\usepackage{geometry}                		
\usepackage{lineno}
\usepackage{abstract}
\usepackage{graphicx}				
\usepackage{float}
\usepackage{sidecap}
\usepackage{natbib}
\usepackage{amssymb,amsmath}
\usepackage{color}
		
\newcommand{\qq}{\mathbf{q}}

\newcommand{\kk}{\mathbf{k}}

\author{Andrij Vasylenko}
\affiliation{Department of Physics, University of Warwick, CV4 7AL, United Kingdom}
\email{a.vasylenko@warwick.ac.uk}
\author{Jamie Wynn}
\affiliation{Cavendish Laboratory, University of Cambridge, CB3 0HE, United Kingdom}
\author{Paulo Medeiros}
\affiliation{Cavendish Laboratory, University of Cambridge, CB3 0HE, United Kingdom}
\author{Andrew Morris}
\affiliation{Cavendish Laboratory, University of Cambridge, CB3 0HE, United Kingdom}
\alsoaffiliation{Department of Physics, University of Warwick, CV4 7AL, United Kingdom}
\author{Jeremy Sloan}
\affiliation{Department of Physics, University of Warwick, CV4 7AL, United Kingdom}
\author{David Quigley}
\affiliation{Department of Physics, University of Warwick, CV4 7AL, United Kingdom}

\title{Encapsulated Nanowires: Boosting Electronic Transport in Carbon Nanotubes}

\begin{document}
\maketitle
\small
\begin{abstract}
The electrical conductivity of metallic carbon nanotubes (CNTs) quickly saturates with respect to bias voltage due to scattering from a large population of optical phonons.
Decay of these dominant scatterers in pristine CNTs is too slow to offset an increased generation rate at high voltage bias. We demonstrate from first principles that encapsulation of 1D atomic chains within a single-walled CNT can enhance decay of ``hot'' phonons by providing additional channels for thermalisation.
Pacification of the phonon population growth reduces electrical resistivity of metallic CNTs by 51\% for an example system with encapsulated beryllium. 
\end{abstract}
Carbon Nanotubes (CNTs) are the most promising candidates for nanoelectronics applications due to their excellent electrical conductivity. \cite{in1,in2,in3}. With these properties applied in integrated circuits and in the field effect transistors' gate electrodes metallic CNTs lead the minituarisation race at the nanoscale\cite{Desai}.
However experimental and theoretical studies show that electronic transport in metallic CNTs undergoes a dramatic decrease beyond a bias of $\approx 0.2$\,eV due to scattering of conduction electrons from a population of high frequency phonons \cite{Amer, transexp2, Park, Yao}. 
Under bias voltage, the process of electron scattering excites new phonons into this population an order of magnitude faster than their decay via thermalisation \cite{epcTh}.
This dominance of excitation over deexcitation results in a growing population of athermal ``hot'' phonons and consequently a non-equilibrium phonon distribution \cite{Bushmaker,DaSilva}. 
Under such conditions, these hot phonons constitute the dominant source of electron scattering, and hence resistivity, in metallic CNTs. 
A mechanism for increasing the rate of phonon thermalisation will therefore enhance the electrical performance of CNTs.

The process of phonon deexcitation involves anharmonic phonon-phonon scattering, hence its rate is dependent on the number of available channels for phonon decay. Previously, several possible solutions for introduction of additional thermalisation channels were suggested that considered a supporting substrate or isotopic disorder \cite{Pop, Mann, isodis}.
Also an alternative mechanism for phonon relaxation via phonon-polariton thermal coupling was proposed for calculation of heat dissipation in CNTs/substrate system on the basis of a parametrised tight-binding model \cite{Rotkin}. 

In this letter we show that reduction of the hot phonon population under bias is readily achievable via encapsulation of 1D nanowires in single-walled CNTs (SWCNT). We consider phonon-phonon relaxation from first principles and  demonstrate that an encapsulated one-dimensional crystal creates additional channels for hot phonon thermalisation, increasing the decay rate. This results in a significant improvement of the voltage-current ratio at high bias voltage.
Transport is studied by solving the set of parameter-free Boltzmann transport equations (BTE) for coupled dynamics of electrons and phonons, whilst all relevant scattering rates are calculated \emph{ab initio} with density-functional perturbational theory (DFPT).  
This route to enhanced transport is attractive due to increasingly well-established methods for growth of 1D crystals inside CNTs \cite{ENN, EN1, EN2, EN3, EN5} and assembly of nanowires into integrated devices \cite{Liebe1, Liebe2}. A broad variety of materials have been encapsulated in this manner, allowing for different degrees of phonon-phonon coupling with the encapsulating CNTs. Furthermore, modern experimental techniques of Raman \cite{RamanOron,RamanDavid, Bushmaker} and ultrafast optical spectroscopy \cite{ultrafast} allow for detailed investigation of how the hot phonon distribution evolves in these novel materials.

\begin{figure}[b!]
\includegraphics[width=6cm]{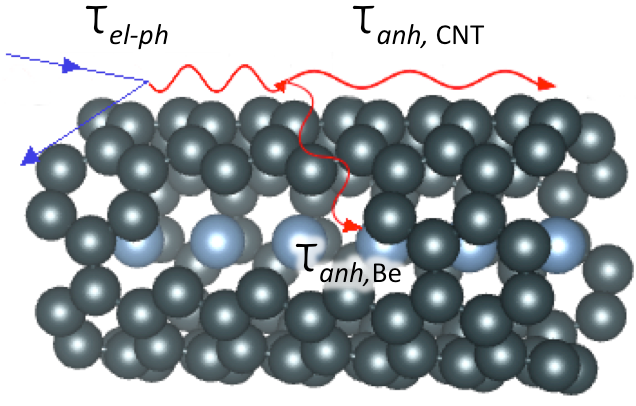}
\caption{
\footnotesize 
Schematic illustration of scattering processes in CNTs. Electrons and phonons are represented by blue and red lines respectively. Here phonon excitation occurs that causes the electron backscattering. The excitation is thermalised due to anharmonic phonon-phonon scattering along two available channels. The characteristic lifetimes of the processes are $\tau_{\text{el-ph}}$ and $\tau_{\text{anh}}$ respectively.}
\label{fig:schematic}
\end{figure}

The competing scattering processes that determine the electrical performance of a material are illustrated in Fig.\ref{fig:schematic}. In metallic CNTs, it has been established that conduction electrons scatter mainly with optical phonons, corresponding to modes at the $\Gamma$ and $K$ points of the graphene Brillouin zone (BZ)  \cite{Bonini}. 
\small
The corresponding electron-phonon scattering times $\tau^{\Gamma,K}_{\text{el-ph}}\sim 0.5, 0.2$\,ps have been previously derived based on DFPT calculations of graphene, and a simple analytical formula accounting for the finite diameter of CNTs \cite{epcTh, Bonini}.  The population of excited phonons grows rapidly in the non-equilibrium regime for voltages higher than $0.2$\,V because thermalisation due to the anharmonic phonon-phonon scattering is significantly slower - from $\sim 3$ps for $\tau^{\Gamma}_{\text{anh}}$ to $\sim 5$ps for $\tau^{K}_{\text{anh}}$ \cite{Bonini}. 

\begin{figure}[b!] 
\includegraphics[width=9.cm]{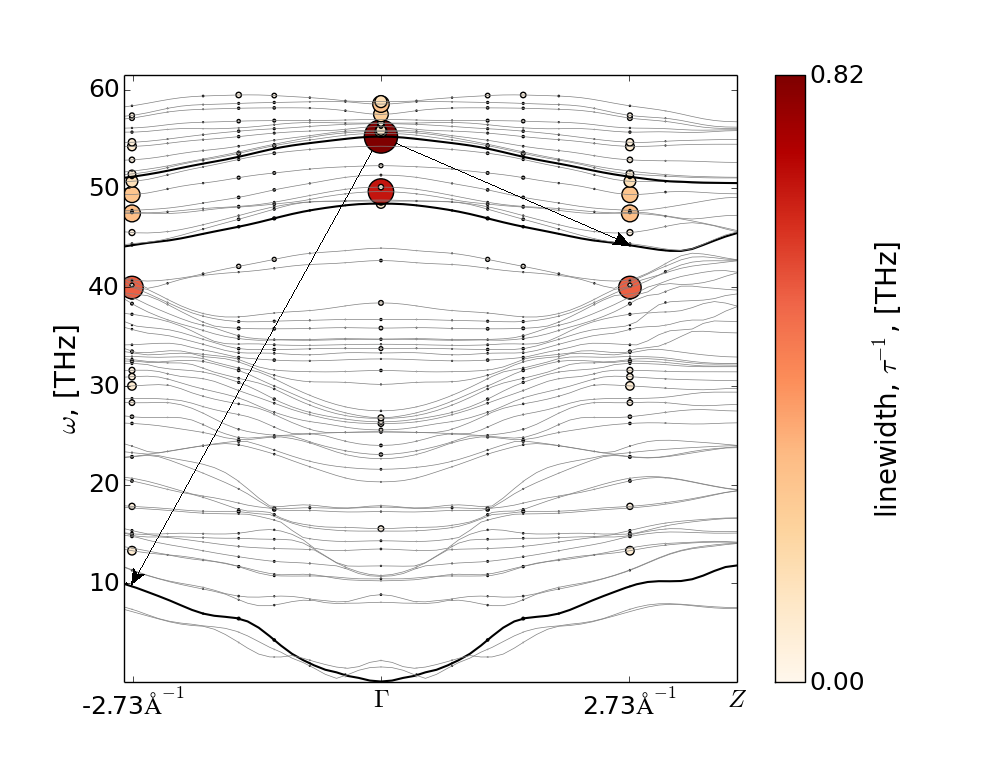}
\caption{
\footnotesize 
Phonon dispersion of SWCNT(6,6) with encapsulated 1D Be atomic chain. The size and colour of the circles relate to the strength of electron-phonon coupling in the Brillouin zone. One of the phonon decay mechanisms available due to Be encapsulation is demonstrated between the highlighted branches: an optical phonon with strong electron-phonon coupling}
\label{fig:dispersion}
\end{figure}
\small
The characteristic lifetimes can be evaluated using Fermi's golden rule \cite{fgr1}. The electron-phonon scattering rate, when only phonon emission is considered (backward scattering) can be expressed as the following:
\begin{equation}\label{elph}
\begin{aligned}
\frac{1}{\tau_{el-ph}^{\nu,\qq}} &= \frac {2\pi} {\hbar} \sum_{i,j,\kk} |g^{\nu}_{ij}(\kk,\qq) |^2 f_{j\kk} (f_{i\kk-\qq}-1) \\
           &\times  (n_{\qq\nu}+1)\delta(\varepsilon_{j\kk-\qq} - \varepsilon_{i\kk} + \hbar\omega_{\nu}(\qq)),
\end{aligned}
\end{equation}
where $g_{ij}^{\nu}(\kk,\qq)$ is the matrix element of electron-phonon interaction between a phonon with mode index $\nu$, momentum $\hbar\qq$ and frequency $\omega$, and electrons in states $i,j$. $f_{j\kk}$ and $n_{\qq\nu}$ are the distribution functions of electrons and phonons respectively. The matrix element $g_{ij}^{\nu}(\kk,\qq)$ is the inner product of electron states and the variation of the electron self-consistent potential with respect to the atomic disturbance due to phonon propagation that we calculate \emph{ab initio} with DFPT \cite{Baroni}.

We consider a periodic system consisting of a metallic single-walled CNT(6,6) (SWCNT) encapsulating a 1D chain of Be - Be@CNT(6,6). It is worth noting that although most materials might be expected to form 1D atomic chains in confinement at small diameters \cite{ENN}, more complex structures become energetically favourable in SWCNTs with diameters larger than $\sim9$ \AA$\,$\cite{EN1, EN2, EN3}. 
We choose Be for this proof-of-concept study as it possesses a bulk lattice parameter commensurate with armchair CNTs, suggesting that effects of artificial mismatch strain will be small. Evidently, phonon lifetimes as well as transport properties of low dimensional carbon materials can change under strain \cite{ETNANO, StrainGraph}, however the viability of the decay mechanism demonstrated in the following is not critically sensitive to such details.

Phonon dispersion of this combined Be@CNT system, excitation of the phonon modes due to the electron-phonon coupling and the corresponding linewidths $1/\tau^{\nu,\qq}$ are presented on the Fig \ref{fig:dispersion}. In agreement with previous calculations \cite{epcTh,Bonini}, the electron-phonon coupling is most pronounced for optical modes at the points corresponding to $\Gamma$ and $K$ of graphene BZ. This does not change significantly upon Be encapsulation.

For the purposes of computing scattering parameters for use within the BTE, coupling was considered for all bands and no values of the considered grid were neglected during integration of the BZ \cite{SI1}.

The excited phonon in a mode $\nu$ with momentum $\textbf{q}_0$ decays into a pair of phonons with opposite momenta, $\pm \bf{q}$ with a rate that can be expressed by the following
\begin{equation}\label{phph}
\begin{aligned}
\frac{1}{\tau_{anh}^{\nu,\bf{q}_0}} & = \frac{\pi\hbar}{16N_q^3m_0\omega_{\nu}(\qq_0)}\sum_{\qq, i,j}  \sum_{\nu', \nu''}|\partial^3_{\qq_0,\qq,-\qq} \mathcal{E} |^2 \\
&\times\frac{\delta(\omega_{\nu}(\qq_0) - \omega_{\nu'}(\qq)- \omega_{\nu''}(-\qq))} {m_i m_j \omega_{\nu'}(\bf{q})\omega_{\nu''}(-\bf{q})}.
\end{aligned}
\end{equation}
The third derivative of the total energy with respect to the atomic displacements, $\partial^3_{\qq_0,\qq,-\qq}\mathcal{E}$, as well as the corresponding phonon frequencies $\omega_{\nu}(\bf{q})$ can be obtained from DFPT \cite{SI1}.

The mechanism of the phonon thermalisation is based on decay into lower energy modes. Thus, by providing additional channels of deexcitation, in our case, phonon modes of an encapsulated nanowire, one can decrease thermalisation time and reduce the hot phonon population. Most channels available for decay are not coupled directly with conduction electrons and therefore do not participate in electrons backscattering. By extension, these do not affect electronic transport. Two of these channels are indicated on the Fig \ref{fig:dispersion}. Frequencies of the marked channels satisfy the summation rule in three-phonon scattering, while conserving energy \cite{Mingo}. Provided that summation rule holds as enforced by $\delta$-function in (\ref{phph}) every atom of the encapsulated nanowire creates three additional vibrational modes with a scattering time $\tau^\nu_{anh,Be}$ and increases the thermalisation rate as ${\tau^{\nu}_{anh}}^{-1} = {\tau^{\nu}_{anh, CNT}}^{-1} + {\tau^{\nu}_{anh,Be}}^{-1}$. 

To evaluate the effect of the reduction in thermalisation time due to Be encapsulation on the high-bias electronic transport in CNTs, we employ the BTE, which involves scattering rates explicitly. 
In this manner the study combines first principles calculations and model kinetic treatment of electronic transport. We emphasise that no other approximations except those inherent to the DFPT and BTE approaches were employed. We consider the all-bands relaxation time approximation for the scattering process, via the calculated characteristic linewidth of each band. 
The initial distribution of electrons is considered equilibrium $f_{j\kk}|_{eq} =(1+\mathrm{e}^{E_j(\kk)/kT})^{-1}$ and evolves according to the following:
\begin{equation}\label{a}
\frac{\partial f_{j\kk}}{\partial t} + \frac{\partial f_{j\kk}}{\partial x}v_j(\kk) +  \frac{\partial f_{j\kk}}{\partial \kk}\frac{eV}{\hbar L} = \frac{f_{j\kk} - f_{j\kk}|_{eq}}{\tau_{el-ph}^{\nu,\qq}},
\end{equation}
where $v_j(\kk)$ is electron velocity of a band $j$ at a point $\kk$ in the BZ, and $V$ is an applied voltage. Electric field $V/L$ is initially uniform along the CNT length $L$ and changes according to the Poisson equation as charge flows. Scattering rates due to the electron-phonon coupling are calculated according to (\ref{elph}) and depend on phonon population. The latter evolves from equilibrium $n_{\qq\nu}|_{eq}= (\mathrm{e}^{\hbar\omega^{\nu}_{\qq}/kT}-1)^{-1}$ during electronic transport as more phonons are excited with increase of voltage bias, while only a fraction of them can be thermalised due to phonon-phonon scattering along the CNT. In order to describe evolution of the phonon distribution we consider the corresponding BTE:
\begin{equation}\label{b}
\begin{aligned}
\frac{\partial n_{\qq\nu}}{\partial t} + \frac{\partial n_{\qq\nu}}{\partial x} v^{\nu}_{ph}(\qq) = &\frac{n_{\qq\nu}-n_{\qq\nu}|_{eq}}{\tau_{el-ph}^{\nu,\qq}} \\ - & \frac{n_{\qq\nu}-n_{\qq\nu}|_{eq}}{\tau_{anh}^{\nu,\qq}},
\end{aligned}
\end{equation}
where $v^{\nu}_{ph}$ is a phonon velocity in a band $\nu$, a collision term describes interplay between excitation and thermalisation of phonons.
The set of equations (\ref{a}),(\ref{b}) is solved consistently by numerical integration in time\cite{SI2}, where all the parameters of the equations are obtained \emph{ab initio} by means of density-functional theory and DFPT as above. The present approach has similar philosophy with that in Ref.\cite{coupled} with several differences in analytical and technical implementation. First, scattering rates are considered explicitly for all bands that evolve due to the change of phonon population. Second, the phonon distribution evolves starting from a room temperature equilibrium distribution rather than zero population. Lastly, a second order upwind discretisation scheme \cite{upwind} is elaborated that makes evolution of distribution functions dependent on electron and phonon velocities for every band $\nu$ and wave vector $\qq$ \cite{SI2}.

\begin{figure}[b!]
\includegraphics[width=9cm]{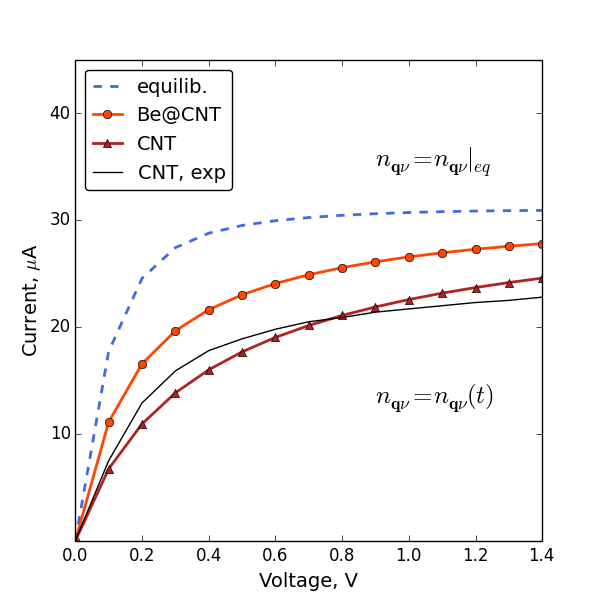}
\caption{
\footnotesize 
$IV$ curves for SWCNT(6,6) and SWCNT(6,6) with encapsulated 1D Be nanowire, $L=300$ nm, $T=300 K$. The dashed blue line is obtained for CNTs with electron-phonon scattering, when phonon distribution does not evolve in time and remains equilibrium. The bottom and top red lines correspond to SWCNT(6,6) and Be@CNT(6,6) respectively with the electron-phonon scattering and anharmonic phonon-phonon scattering taken into account. The black line represents experimental values from \cite{Park}.}
\label{fig:ivcurve}
\end{figure} 
\small
The significance of effect of electron-phonon and phonon-phonon scattering on electronic transport in CNT can be understood from Fig.\ref{fig:ivcurve}. When phonon population is neglected $n_{\qq\nu} = 0$, electric current shows ballistic behaviour, with continuous quasi-linear growth with voltage increase (not shown on the Fig.\ref{fig:ivcurve}). We demonstrate that electron-phonon scattering with a static phonon population has a major impact on the resistivity of CNTs, causing electric current saturation. However, optimal agreement with experiment is achieved when the phonon population is allowed to evolve away from equilibrium during the electron flow. Evidently, accumulation of athermal optical phonons contributes to further saturation of electric current. The Joule loss in the systems thermally coupled to a substrate within a simple model proposed in \cite{Freitag} has additional effect on the saturation behaviour, which however is less pronounced in comparison to the contribution from electron-phonon scattering \cite{SI3}. Bader population analysis \cite{bader} demonstrates no charge transfer between Be nanowire and SWCNTs upon encapsulation, thus there is no increase of the number of carriers in the composite system\cite{SI5}. The electronic band structure of Be@CNT(6,6) as well as velocities of electrons do not differ considerably from those of a pristine SWCNT. Moreover calculations on an isolated 1D chain of Be atoms demonstrate that the wire itself does not conduct electrons \cite{SI4}. Nevertheless the presence of a Be nanowire encapsulated within the CNT pacifies the growth in hot phonons population and results in a reduction of electrical resistivity of SWCNTs in room temperature by 51.2\% for a threshold voltage of 0.2V and by 19.8\% for 1V bias. 

The considered model kinetic behaviour is justified for the systems, where electronic scattering by lattice vibrations is dominant in the whole range of temperatures as no scattering with impurities is taken into account. This assumption allows us both to consider temperature effects in terms of distribution functions only, and also to employ the zero temperature scattering probabilities. Description of elastic scattering with defects or temperature dependence of the coupling constants would require an approach beyond the relaxation time approximation that would consider such effects of many-phonon scattering as second sound \cite{secondsound}.

In conclusion, we confirm that generation of hot phonons saturates electric current in metallic CNTs under high-voltage bias. We demonstrate that encapsulation of Be atomic chains within a SWCNT mitigates this effect. From first principles it is calculated that encapsulated Be nanowire reduces thermalisation time of excited optical phonons. The occupation of optical phonons and its evolution is experimentally accessible by ultrafast optical and Raman spectroscopy and its time-resolved modification\cite{Bushmaker, ultrafast}. It is established that electrical conductivity of metallic SWCNT is in great part determined by the strength of electron-phonon coupling. It is worth noting that the later depends effectively on electron and phonon distributions that in general have strong temperature dependence \cite{SusGraph}. A \emph{thermally} conductive encapsulated nanowire can play the role of a heat sink that moderates increase of temperature during electronic transport. The degree to which an encapsulated atomic chain affects electrical resistivity also depends on the phonon dispersion characteristic of a particular nanowire, i.e. on the number of modes that satisfy conservation of energy and phonon selection rules during decay \cite{Mingo}. We can expect that the appropriate choice of material, for encapsulation as well as the relevant structure of the nanowire, could lead to a more pronounced effect of boosting the electronic transport in SWCNTs in wide range of voltage bias.

\section*{Acknowledgements}
We acknowledge financial support from the EPSRC (Grant Numbers EP/M010643/1 and EP/M011925/1).
This work used the ARCHER UK National Supercomputing Service \newline (http://www.archer.ac.uk) and facilities provided by the Research Technology Platform for Scientific Computing, University of Warwick.
\bibliography{myref}{}

\onecolumn
\clearpage
\pagebreak
\setcounter{equation}{0}
\setcounter{figure}{0}
\setcounter{table}{0}
\setcounter{page}{1}
\makeatletter
\renewcommand{\theequation}{S\arabic{equation}}
\renewcommand{\thefigure}{S\arabic{figure}}
\renewcommand{\bibnumfmt}[1]{[S#1]}
\renewcommand{\citenumfont}[1]{S#1}
\section*{Supplemental Material. Encapsulated Nanowires: Boosting Electronic Transport in Carbon Nanotubes}
\begin{figure*}
\includegraphics[width=10cm]{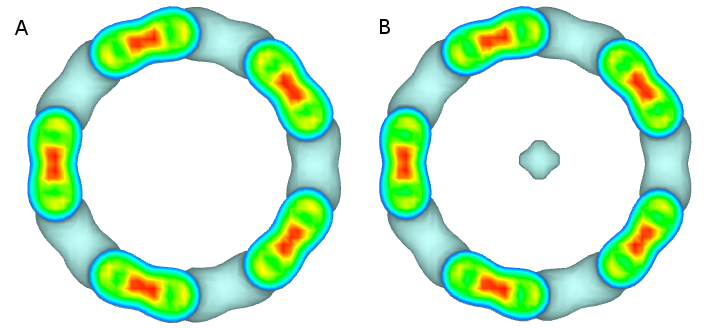}
\caption{
\footnotesize
Isosurface of electronic charge density (a slice across the axes): (A) pristine SWCNT; (B) Be@SWCNT}
\label{fig:iso}
\end{figure*}

\small
\textit{Ab initio calculations.}
Electron-phonon and phonon-phonon scattering rates were calculated with DFPT as it is implemented in Quantum Espresso suite \cite{QE1}. We used norm-conserving pseudopotentials (as required for the response calculations with respect to atomic displacement) within Becke-Lee-Yann-Park approach \cite{Becke1} for calculation of non relativistic exchange-correlation interaction. The unit cell $30\times 30 \times 2.46$\,\AA\, consists of a fraction of SWCNT(6,6) of 2.46\,\AA\, long surrounded by vacuum in non-axial directions. Electron occupation numbers were considered in Fermi-Dirac statistics with 0.025 Hartree broadening.  For calculations of electron band structure, electron velocities and electron-phonon constants BZ was sampled with $1\times1\times500$ grid in Monkhorst-Pack scheme that yields $\frac{2\pi}{251\times2.46}$$\AA^{-1}$ spacing in BZ. The corresponding BZ sampling in calculations of phonon dispersion, phonon velocities and anharmonic phonon coupling used a coarse grid $1\times1\times40$ was used that was further Fourier-interpolated to a $1\times1\times200$ fine grid. With a plane-wave-cutoff energy of 60 Hartree, phonon calculations converged within 0.5 cm$^{-1}$ for optical phonons at $\Gamma$ point.
\begin{figure*}[h!]
\includegraphics[width=\textwidth]{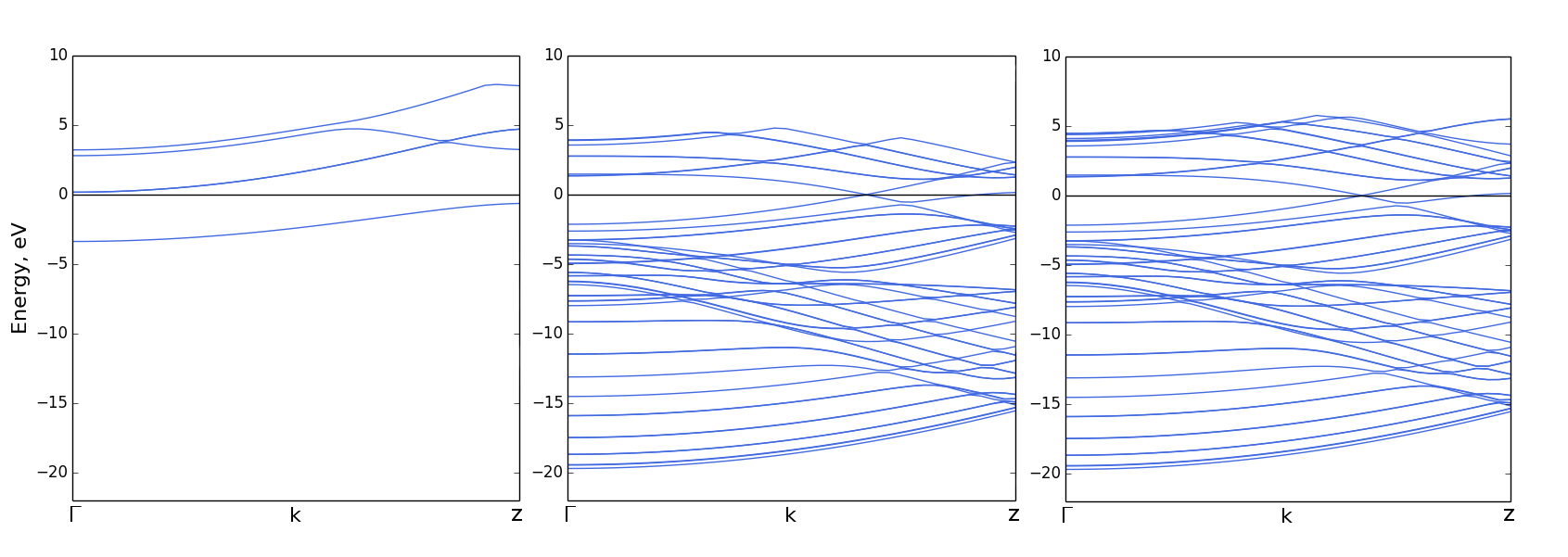}
\caption{
\footnotesize
Band structures of the combined and isolated systems. Fermi level is aligned at 0 eV, $\kk$ indicates axial direction in BZ. \textit{Left:} Isolated 1D Be chain. Wide band gap indicates that 1D Be atomic chain does not transport electrons when isolated. \textit{Centre:} Pristine SWCNT(6,6). Dirac cone at Fermi level indicates metallic nature of a tube. \textit{Right:} Band structure of Be@CNT(6,6) alters insignificantly from the one of pristine CNTs in terms of electron mobility and available energy levels.}
\label{fig:bs}
\end{figure*}
\newline
\renewcommand{\citenumfont}[1]{#1}
\textit{Charge analysis.} We analyse the two systems, pristine CNTs and CNTs with encapsulated Be, on the basis of the calculations conducted above. Visualised isosurfaces of the electronic charge density of the systems are presented on the Fig.\ref{fig:iso} and demonstrate no apparent changes inthe electronic structure of CNTs, which may be credited to the chemical inertness of the latter. Further analysis conducted with the Bader method \cite{bader} shows redistribution of the charge upon geometry optimisation of CNTs after Be atom is encapsulated (See Table \ref{table:bader}). The change of the total charge on the CNT's walls due to Be encapsulation is negligible: the resulting increase of 0.0301$e$ is comparable to the charge of surrounding vacuum in the unit cell and is subject to definition of the atomic volume exploited in the approach.
\renewcommand{\citenumfont}[1]{S#1}
\begin{table}
\caption{
\footnotesize
Bader charge analysis of the pristine SWCNT and Be@SWCNT.}
\label{table:bader}
\footnotesize
\begin{tabular} { p{2.5cm} p{2cm} p{2cm} p{2cm} c}
\hline
 & SWCNT & Be@SWCNT & Be(embeded) & $\Delta e$ \\
\hline                                      
Total charge, $e$ & 80.1396 & 84.1499 & 3.9725 & 0.0301 \\
\hline
\end{tabular}
\end{table}

\begin{figure*}
\includegraphics[width=12cm]{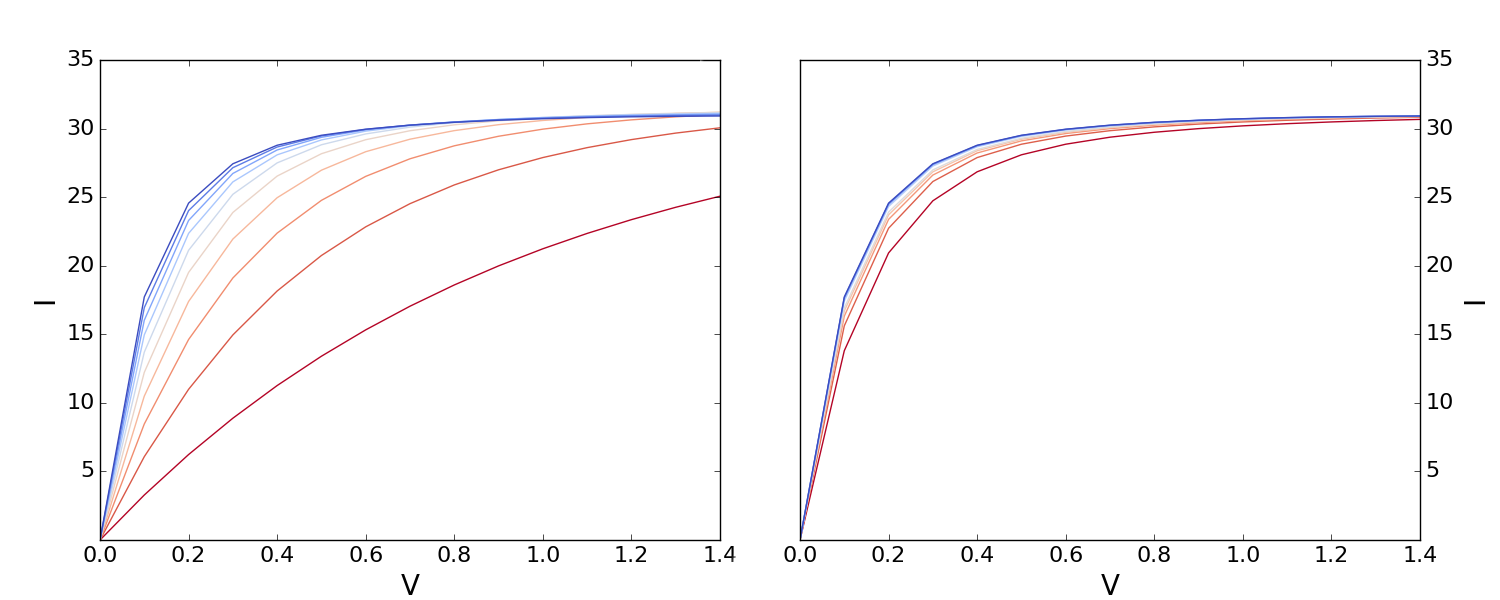}
\caption{
\footnotesize
\textit{Left:} Convergence of the IV curve with time. Steady state (top blue curve) is reached when $t = 3$ps. \textit{Right:} Convergence with respect to the sparsity of the space grid: $\Delta x$ from 50nm (bottom curve) to 1nm (top curve).}
\label{fig:ivconv}
\end{figure*}

\newpage
\textit{Integration of BTE.} For numerical integration of the BTE we employed a second-order upwind discretisation scheme, in which in the left-hand side of the Eq.(\ref{a}) the electron distribution function $f^j_{x,k,t}$ on every band $j$ changes in time, on direct and reciprocal space grids
\begin{equation}\label{eq:s1}
f^j_{x,k,t+1} = f^j_{x,k,t}-\Delta t (v^{\pm}_jf^{\pm}_x + F^{\pm}f^{\pm}_k),
\end{equation}
where
\begin{equation}\label{eq:s2}
\begin{aligned}
f^{+}_{\alpha} = \frac{3f_{\alpha} - 4f_{\alpha-1} + f_{\alpha-2}}{2\Delta\alpha},\\
f^{-} = \frac{-f_{\alpha+2}^n + 4f^n_{\alpha+1} -3f^{n}_{\alpha}}{2\Delta\alpha},
\end{aligned}
\end{equation}
where $\alpha={x,k}$, the sign of $f_{\alpha}^{\pm}$ depends on the sign of electron velocity $v^{\pm}$ (electric field $F^{\pm}=\Delta\varphi(x,t)/\Delta x$) at the particular grid point. Electric field is updated with the change of charge $\rho(x,t)$ and electrostatic potential $\varphi(x,t)$ according to the Poisson-Boltzmann equation 
\begin{equation}
\hat{L}\cdot\varphi(x,t) = -\rho(x,t) - \int f^{j}_{x,k,t}dk,
\end{equation}
where Laplacian operator $\hat{L}$ for the steady state is represented with discrete Laplacian matrix. The phonon distribution function $n^{\nu}_{x,k,t}$ evolves similarly to (\ref{eq:s1}),(\ref{eq:s2}) with the exclusion of the terms dependent on electric field.
The distribution functions $f^j_{x,k,t}$ and $n^{\nu}_{x,k,t}$ are sampled on a 3-dimensional grid $N_b\times300\times200$, each dimension corresponding to a band, direct and inverse space coordinates respectively. Starting from equilibrium values both distributions as well as charge and corresponding electrical potential are updated each 1 fs for 3000 time steps. The convergence of the BTE solution when electron-phonon scattering is taken into account (that corresponds to the light blue line on Fig.\ref{fig:ivcurve}) with discretisation parameters, $t$ and $\Delta x$ is presented in Fig.\ref{fig:ivconv}.

\begin{figure}[h!]
\includegraphics[width=9.cm]{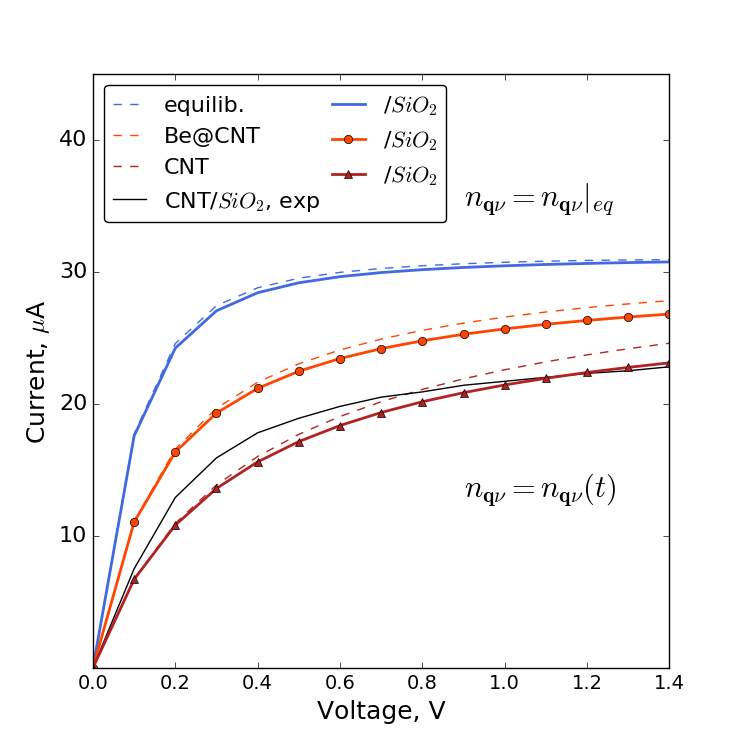}
\caption{
\footnotesize
Dashed lines correspond to the current saturation in the systems not coupled thermally with a substrate, $T_0$=300K (reproducing the lines on the Fig.3 in the main text). The IV curves during self-heating of CNTs on a thermally coupled $SiO_2$ substrate are presented with solid lines. The black line represents experimental value of current in CNT(6,6) on a $SiO_2$ substrate.}
\label{fig:satur}
\end{figure}

\renewcommand{\citenumfont}[1]{#1}
\textit{Self-heating effect on electronic current in CNTs on a thermally coupled substrate.} Here we consider the effect of coupling CNTs to a thermal bath representing a polar substrate (following the procedure in \cite{Rotkin}) and demonstrate that the effect on transport is small in comparison to that of additional phonon decay channels resulting from nanowire encapsulation. For a polar substrate, thermal conductivity $\kappa$, the Joule losses can be accounted for via self-consistent simulation of temperature increase as the following:
\begin{equation}\label{heat}
T = T_0 + \frac{I\cdot V}{L\cdot\kappa}.
\end{equation}
When fully coupled with a $SiO_2$ substrate, thermal conductivity $\kappa = 1.4 W/(m\cdot K)$, the current in CNTs undergoes further saturation. In this picture, self-heating and coupling with a substrate to the smaller degree contribute to the electronic current saturation in comparison with the electron-phonon scattering and nonequilibrium phonon population (See Fig.\ref{fig:satur}). 

\end{document}